\documentclass[twocolumn,amsmath,amssymb,floatfix]{revtex4}

\usepackage{bm}
\usepackage{amsmath}
\usepackage{epsf}

\newcommand{\beq}{\begin{equation}}

\newcommand{\eeq}{\end{equation}}
\newcommand{\beqa}{\begin{eqnarray}}
\newcommand{\eeqa}{\end{eqnarray}}

\newcommand{\bn}{\hat{\bf n}}

\def\simlt{\lesssim}
\def\simgt{\gtrsim}

\newcommand{\mnras}{Mon. Not. R. Astron. Soc.}

\begin{document}
\bibliographystyle{apsrev}
\title{Measuring Dark Energy Clustering with CMB-Galaxy Correlations}
\author{Wayne Hu$^{1}$, Ryan Scranton$^{2}$}
\affiliation{
$^{1}$Kavli Institute for Cosmological Physics, Department of Astronomy \& Astrophysics,
Enrico Fermi Institute, University of Chicago, Chicago IL 60637\\
$^{2}$University of Pittsburgh, Department of Physics and Astronomy, Pittsburgh, PA 15260\\}

\begin{abstract}
\baselineskip 11pt
The integrated Sachs-Wolfe (ISW) effect in the cosmic microwave background (CMB)
as measured through its correlation with
galaxies provides a unique opportunity to study the dynamics of the dark energy
through its large scale clustering properties. 
 Ultimately, a deep all-sky
 galaxy survey out to $z\sim 2$ can make a $\sim 10 \sigma$ or $\sim 10\%$ measurement of the correlation
 and limit $\sim 3\%$ changes in the gravitational potential or total density fluctuation
 due to dark energy clustering on
 the Gpc scale.
 A canonical single scalar field or quintessence model predicts that these
clustering effects will appear on the horizon scale with a strength that reflects the evolution
of the dark energy density.  In terms of a constant equation of state, this would allow tests
of the quintessence prediction for models where $|1+w| \simgt  0.05$.
\end{abstract}
\maketitle

\section{Introduction}

The two generic signatures of a dynamical form of the dark energy are an energy
density that evolves as a function of time and one that varies  as 
a function of space.  Clustering in the dark energy is in fact an unavoidable consequence 
of time evolution on superhorizon scales.  
Prospects for measuring the density evolution or equation of state
of the dark energy are well known.    Detecting its spatial clustering is considerably
harder but of at least equal fundamental importance.  In particular, the clustering tests
 the quintessence hypothesis, dark energy as a 
single scalar field with a canonical kinetic term \cite{CalDavSte98,Hu98}.

Perhaps the best hope of studying the clustering of the dark energy lies in its
effect on the cosmic microwave background (CMB).
In general, the effect of having a ``smooth" component
that alters the expansion rate without clustering with the dark matter
is to make gravitational potentials decay in the linear regime.  As a CMB photon
transits a decaying potential well, it picks up a net blueshift.  
When integrated across the potential hills and wells along the line of sight, this
effect leaves a temperature anisotropy across the line of sight
called the 
integrated Sachs Wolfe (ISW) effect  \cite{SacWol67,KofSta85}.  On scales
where the dark energy is clustered, this effect is reduced in a potentially 
measurable way.  It is a curious coincidence that the observed large angle 
temperature anisotropy is also suppressed compared with a cosmological
constant model (e.g. \cite{Benetal03} and \cite{BeaDor04,GorHu04} for a possible ISW connection).  

Since the decay of the gravitational potential is a direct consequence of the dark
energy, the ISW effect is more sensitive to changes in the
clustering of the dark energy than other clustering statistics.  Unfortunately
this sensitivity is hidden in the CMB by the anisotropy  from recombination.
The ISW contributions  can be isolated by cross correlating the temperature
field with other tracers of the gravitational potential, in particular galaxies
 \cite{CriTur96}
and gravitational lensing statistics \cite{GolSpe99,SelZal98,Hu01c}.
 
 Recently the ISW effect has been detected by cross correlating temperature maps
 from WMAP with radio galaxies,
 optical galaxies, infra-red galaxies and X-ray sources 
 \cite{BouCri04,FosGaz04,Scretal04,AfsLohStr04}.  
 The current measurements are not sufficient for much beyond a 
simple detection of the effect, but do indicate that the technical 
issues involved are well within in the capabilities of current and future 
surveys.

  In this paper we study the prospects for constraining the smoothness of the
 dark energy with future galaxy surveys.  This study extends  that of Afshordi \cite{Afs04}
who mainly considered  prospects for constraining the equation of state of the
 dark energy under the small scale Limber approximation.
ISW constraints on a smooth equation of state are not competitive with other probes
of dark energy evolution due to the unavoidable cosmic variance of the CMB temperature
field, whereas they may provide the best means of studying the clustering of the dark energy.

 The outline of the paper is as follows. In \S \ref{sec:clustering}, we review the predictions
 and parameterization of the large scale clustering of the dark energy.  
 In \S \ref{sec:cross}, we calculate the ISW-galaxy correlation under the general
 all-sky framework applicable to correlated galaxy distributions and fine redshift resolution.
 In \S \ref{sec:forecasts} we study the potential of galaxy surveys to constrain the
 clustering of the dark energy.

\section{Dark Energy Clustering}
\label{sec:clustering}

\begin{figure}[t]
\centerline{\epsfxsize=3.5truein\epsffile{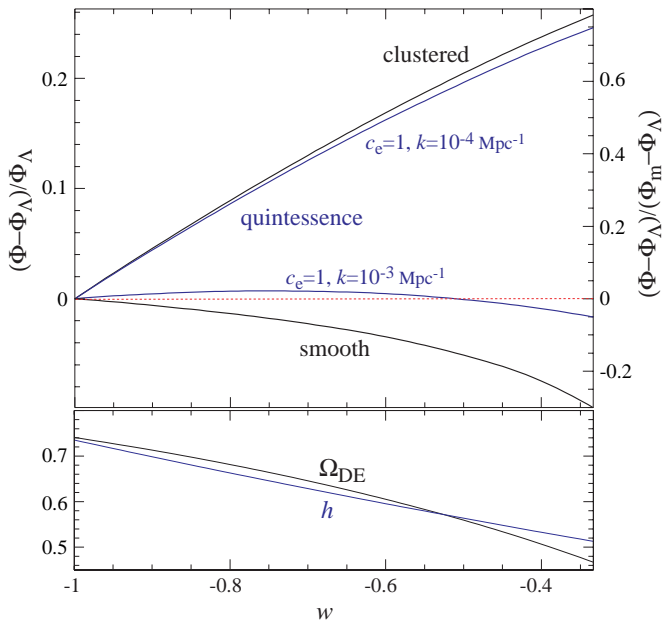}}
\caption{Fractional difference in the gravitational potential today between smooth and clustered
dark energy models as a function of $w$ (upper panel) where $\Omega_{\rm DE}$ and
$h$ are adjusted to keep the distance to recombination and the expansion rate at high-$z$
fixed.  Also shown are the predictions for quintessence dark energy (canonical kinetic
term with sound speed $c_e=1$) near the horizon scale $k=10^{-4}$ Mpc$^{-1}$ and
near the broad peak of the ISW effect at low multipoles $k=10^{-3}$ Mpc$^{-1}$.}
\label{fig:phi}
\end{figure}

With adiabatic initial conditions and in 
the absence of non-gravitational forces,  all energy density components, including
the dark energy, are clustered.  In this limit,
the gravitational potential or Newtonian curvature $\Phi$
evolves in a flat universe according to the relation \cite{Bar80,HuEis99}
\begin{equation}
\Phi  = \left( 1 - {H(a) \over a} \int_{0}^a {da ' \over H(a')} \right) \zeta_i\,,
\label{eqn:clustered}
\end{equation}
where $\zeta_i$ is the initial comoving curvature and $H$ is the Hubble parameter.
  Note that during epochs when the
expansion is dominated by a species with a constant equation of state parameter
$w_{T}=p_{T}/\rho_{T}$, $H(a) \propto a^{-3(1+w_{T})/2}$ and the gravitational potential is constant
\begin{equation}
\Phi = {3(1+w_{T}) \over 5 + 3w_{T}} \zeta_i \,.
\end{equation}
With only gravitational forces, the ISW effect vanishes in a flat, 
constant $w_{T}$ universe  except for the special case of $w_T=-1$ where the constant is zero. 

To accelerate the expansion, dark energy requires relativistic stresses where the pressure
is comparable to the energy density in magnitude.   Stress gradients can prevent the
clustering of the dark energy on small scales.   Given dark energy that is smooth
compared with the dark matter,  the gravitational potential evolves in a flat universe as
\begin{eqnarray}
\frac{d^2 \Phi}{d \ln a^2}  + \left[ \frac{5}{2} - \frac{3}{2} w(a) \Omega_{\rm DE}(a) \right]
\frac{d \Phi}{d \ln a} && \nonumber\\
 + 
\frac{3}{2}[1-w(a)]\Omega_{\rm DE}(a) \Phi &=&0\,.
\label{eqn:smooth}
\end{eqnarray}
Here $w(a)= p_{\rm DE}/\rho_{\rm DE}$ is the equation of state for the 
dark energy and $\Omega_{\rm DE}(a) = \rho_{\rm DE}(a)/\rho_{\rm crit}(a)$.  Where
no argument is given $a=1$ is assumed and the Hubble constant
 $H_0 = H(a=1) = 100h\,$km s$^{-1}$ Mpc$^{-1}$.
The initial conditions for scales well above the horizon at matter-radiation equality
are provided by Eq.~(\ref{eqn:clustered}) with a starting
epoch in the matter-dominated limit.   For smaller scales, the initial conditions are modified
by the usual transfer function to account for radiation stresses.

In Fig.~\ref{fig:phi} (upper panel), we show the fractional difference in the gravitational potential today for
the smooth and clustered regime
relative to a cosmological constant $w=-1$ model.  In this model $\Phi$ has decayed from 
$\Phi_m=3 \zeta_i /5$ in the matter dominated $w_T=0$ regime
 to $\Phi_\Lambda = 0.75 \Phi_m$  by the present.  
 
On the right axis, we normalize
the curves relative to the net change in the gravitational potential $\Phi_m - \Phi_\Lambda =
0.25 \Phi_m$ in the $w=-1$ model.
Since the ISW effect is sensitive to the change in the gravitational potential, this factor of 3 enhancement of the difference reflects the observational effect.  This enhancement is the reason why the
ISW effect is more sensitive to dark energy clustering than other measures of clustering.

Note that here and throughout when considering models of different $w$,
$\Omega_{\rm DE}$ and $h$ are 
adjusted to keep the well-constrained distance to recombination and
high redshift expansion rate fixed (Fig~\ref{fig:phi}, lower panel).    
This adjustment guarantees that the CMB
 predictions for the various dark energy models
will be indistinguishable aside from the ISW effect of interest (see also Fig.~\ref{fig:cl}).

\begin{figure}[t]
\centerline{\epsfxsize=3.5truein\epsffile{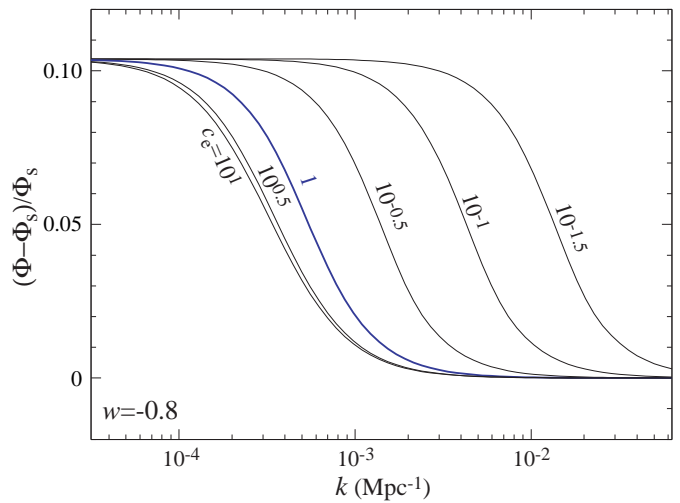}}
\caption{Fractional change in the gravitational potential in the transition regime
between smooth ($\Phi_{s}$) and clustered dark energy for several choices of the sound speed and
a $w=-0.8$ model.
The quintessence prediction corresponds to $c_e=1$.}
\label{fig:csphi}
\end{figure}

The transition between the smooth and clustered
regimes depends on the physical model for the dark energy.    It is usefully parameterized by the 
effective
sound speed of the dark energy $c_e$, defined through 
\begin{equation}
c_e^2= {\delta p_{\rm DE} \over \delta \rho_{\rm DE}}
\end{equation}
in a ``rest frame" coordinate system where the momentum density of the dark energy vanishes
\cite{Hu98}.   
In terms of the gravitational potential phenomenology, the minimum of the sound horizon
and (particle) horizon separates the smooth and clustered regime.   Above the horizon,
there is no unique definition of smoothness for a dark energy component whose density
varies with time given the freedom to choose the time slicing in general relativity \cite{CalDavSte98}.

In a scalar field dark energy model, the sound speed so defined is exactly given
in linear theory
by the form of the kinetic energy as a function of the field \cite{GarMuk99,DedCalSte03}.  
For
a canonical kinetic term, the sound speed $c_e=1$.   As a dark energy candidate,
a canonical scalar field is dubbed quintessence and one with an alternative
kinetic term k-essence \cite{ArmMukSte00}.
For
quintessence, 
the transition between the smooth and clustered regimes
thus occurs near the horizon scale.  In Fig.~\ref{fig:phi}, we also show the gravitational potential
amplitude for a scalar field model at $k=10^{-3}$ Mpc$^{-1}$ and $k=10^{-4}$ Mpc$^{-1}$.
Note that the horizon scale
\begin{equation}
\eta_0 = \int_0^1 { dt \over a } \approx 14 {\rm  Gpc}\,,
\end{equation}
 and so the horizon wavenumber
$k_H\equiv \eta_{0}^{-1} \approx  7 \times 10^{-5}$ Mpc$^{-1}$.  
In the $w=-1$ model contributions to the ISW effect at the scale of the
quadrupole are broadly distributed
around $k=10^{-3}$ Mpc$^{-1}$ due to projection effects (see e.g. \cite{GorHu04}).  Thus a 
fair fraction of the contributions to the ISW effect will come from the transition regime if
$w \ne -1$ (e.g. \cite{Hu98,Hu01c,WelLew03}).  

In Fig.~\ref{fig:csphi}, we
show the effect of dark energy clustering as a function of scale for $c_e= 10^{-1.5}, 10^{-1} \ldots
10^1$ for $w=-0.8$ the smooth limit $\Phi_s$ of Eq.~(\ref{eqn:smooth}).  We employ these numerical
results for the potential spectrum at the present in the following sections.  For the
time evolution employed in the next section, we use 
the interpolation between solutions of the clustered and smooth regimes in 
Eq.~(\ref{eqn:clustered}) and (\ref{eqn:smooth}) given in \cite{Hu01c},  valid for $c_{e}
\le 1$.

Note that the clustering effects saturate close to $c_e=1$.
For this reason, we will take models with $c_e \le 1$.  For illustrative purposes
we will also typically employ the $w=-0.8$ model but discuss the conversion of constraints
to general dark energy models in \S \ref{sec:forecasts}.  
\section{ISW Cross-Correlation}
\label{sec:cross}
The effects of dark energy clustering can be seen on the two-point correlation between
observed fields that depend on  the gravitational potential
such as the CMB temperature and the galaxy number density.
Let us assume that
a field $x(\bn)$ as a function of the angular position $\bn$ on the
sky is a weighted projection of the potential field $\Phi({\bf x};z)$
\begin{equation}
x(\bn) = \int dz W^x(D\bn;z) \Phi(D\bn;z)\,,
\end{equation}
where the weight $W^x$ can include differential operators acting on the field.   Here
$D= \int_0^z dz/H(a)$ is the comoving distance to redshift $z$.
Then the
angular cross correlation between two observed fields $x$ and $x'$ is given by \begin{equation}
\langle x(\bn) x'(\bn') \rangle = \sum_\ell  P_\ell(\bn\cdot\bn')  {{2\ell+1} \over 4\pi}C_\ell^{xx'}\,,
\end{equation}
where $P_\ell$ is the Legendre polynomial and $C_\ell^{xx'}$ is the
cross power spectrum.  The angular power spectrum itself is given
by
\begin{equation}
C_\ell^{xx'} = 4\pi \int {d^3 k \over (2\pi)^3} I_\ell^x(k) I_\ell^{x'}(k) P^{\Phi\Phi}(k;0)\,,
\end{equation}
where $P^{\Phi\Phi}(k;z)$ is the 3D power spectrum of the potential field
\begin{equation}
\langle \Phi({\bf k};0) \Phi({\bf k}';0) \rangle = (2\pi)^3 \delta({\bf k}-{\bf k}') P^{\Phi\Phi}(k;0)
\end{equation}
and the weights
\begin{equation}
I_\ell^{x}(k) = \int dz {\Phi(k;z) \over \Phi(k;0)} W^x(k;z) j_\ell (k D) \,.
\end{equation}
For the ISW effect
\begin{equation}
W^{I}(k;z) = 2{\partial \ln \Phi \over \partial z} \,.
\end{equation}
Note that in spite of the fact that in reasonable dark energy models the gravitational
potential changes most rapidly around the current epoch, the local contributions
to the angular power
spectrum are strongly suppressed by the $j_\ell$ projection 
factors. The contributions from finite $k$  locally 
appear in the monopole moment of the temperature field which makes a small and inseparable
contribution to the background temperature.

\begin{figure}[t]
\centerline{\epsfxsize=3.5truein\epsffile{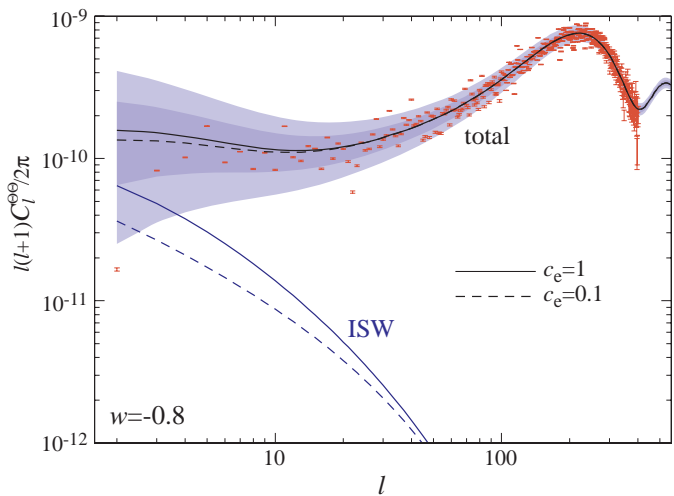}}
\caption{CMB temperature power spectrum for the $w=-0.8$ model compared
with the WMAP 1 year data with noise errors and the $68\%$ and $95\%$
cosmic variance confidence bands plotted for $c_e=1$.  The $c_e=0.1$ model
is difficult to distinguish from the $c_e=1$ model even with perfect data despite
the fact that the ISW effect changes by up to a factor of 2.}
\label{fig:cl}
\end{figure}

In Fig.~\ref{fig:cl} we show CMB temperature power spectra 
for the $w=-0.8$ ($\Omega_{\rm DE}=0.68$) 
model with the ISW contribution separated out for $c_e=1$ and $c_e=0.1$.  
In addition to the dark energy parameters, we will assume a physical non-relativistic 
matter
density of 
$\Omega_m h^2 = 0.14$, a physical baryon density of $\Omega_b h^2=0.024$, a reionization optical depth of  $\tau=0.17$, an initial scale invariant comoving curvature spectrum with amplitude 
$\delta_\zeta = 5.07 \times
10^{-5}$ (WMAP $A=0.87$ \cite{Speetal03}) and slope $n = 1$.

Although the ISW contributions for the two models in Fig.~\ref{fig:cl} 
differ by up to a factor of 2 as is
consistent with Fig.~\ref{fig:phi}, the difference in the total temperature field is
considerably less due to contributions from other effects.  Given the
cosmic variance of the temperature field also shown, the subtle difference
is difficult to detect from the CMB alone.  

The intrinsic sensitivity of the ISW effect to dark energy clustering 
can be extracted from 
cross correlations with the galaxy number density fluctuations. 
For the galaxy clustering, there is a number density fluctuation field for
each population of galaxies.  We denote the $i$th population as having a
number density fluctuation $g_{i}$.   The weights are given by 
\begin{equation}
W^{g_i}(k;z) = {n^{g_i}(z) \over n_{A}^i} { 2\over 3}\left( {k\over H_0} \right)^2 {a \over \Omega_m} b^{g_i}(k;z)\,,
\end{equation}
where the $n^{g_{i}}(z)$ is the redshift distribution of the population with an angular
number density of  $\int dz\, n^{g_i}(z)=n_A^i$
and $b^{g_i}(k;z)$ is the galaxy bias of the population.  
For the galaxy bias we take the parameterized halo 
model described in \cite{HuJai03}.  As long as galaxy number fluctuations trace the total
density fluctuation or gravitational potential in the linear regime, the details of this model 
do not affect the end results.  This is because in the linear regime, the galaxy auto correlation $C_\ell^{g_{i} g_{j}}$
determines the bias parameters for the ISW-galaxy cross correlation.  Marginalization of 
halo parameters mainly eliminates the small amount of information coming 
from the non-linear regime.

\begin{figure}[t]
\centerline{\epsfxsize=3.1truein\epsffile{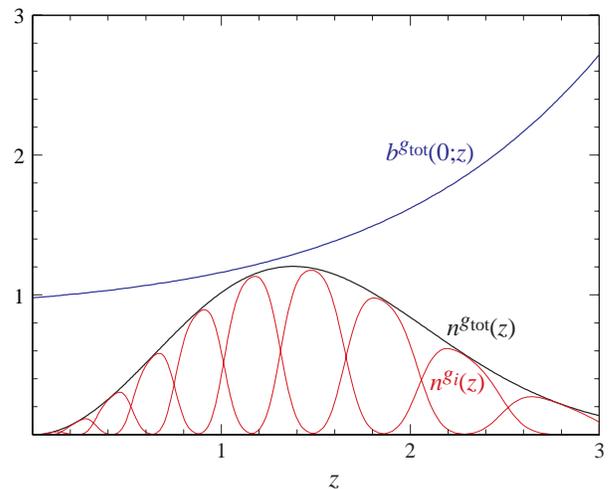}}
\caption{Bias and redshift distribution of the galaxies.  The total number density $n^{g_{\rm tot}}$ 
(here multiplied by  $2/n_A^{g_{\rm tot}}$, with $n_A= 70$ gal arcmin$^{-2}$ for clarity) 
defines
the linear bias $b^{n_{\rm tot}}(0;z)$ under the halo model \cite{HuJai03}.   This total number
density is divided into photometric redshift bins of $5 \sigma(z)$ with
photometric redshift errors of $\sigma(z)=0.03 (1+z)$.}
\label{fig:redshift}
\end{figure}

 For illustrative purposes, let us 
take populations to be selected from an overall redshift distribution of the form
\begin{equation}
n^{g_{\rm tot}}(z) \propto z^2 e^{-(z/z_n)^2}\,,
\end{equation}
with $z_n$ adjusted to give a median redshift of $z=1.5$ and a total
galaxy number density of $n_A^{\rm tot} = 70$ gal/arcmin$^2$.   Under the halo model of
\cite{HuJai03} this number density determines the linear bias shown in Fig.~\ref{fig:redshift}.
 These specifications
are close to what can be achieved from an LSST type survey \cite{LSST}.
This parent distribution can be subdivided into multiple populations  
\begin{equation}
n^{g_{\rm tot}}(z) = \sum_i n^{g_i}(z)
\end{equation}
through photometric
redshifts.  To approximate the redshift binning, let us suppose that photometric
redshift estimates are distributed as a Gaussian with an rms fluctuation of $\sigma(z)$.
A top hat cut in photometric redshift then becomes smooth overlapping distributions in
actual redshift
\begin{equation}
n^{g_i}(z) = {1 \over 2} n^{g_{\rm tot}} (z)\left[
 {\rm erfc}\left( {{z_{i-1} - z} \over \sqrt{2} \sigma(z) } \right)
-  {\rm erfc}\left( {{z_{i} - z} \over \sqrt{2} \sigma(z) } \right)\right] \,,
\end{equation}
where erfc is the complementary error function.
We take as a fiducial model $\sigma(z) = 0.03 (1+z)$ and choose the binning 
to span $5\sigma(z)$ for  a total of 5 bins out to $z=1$ and 10 bins out to $z=3$
(see Fig.~\ref{fig:redshift}).

\begin{figure}[t]
\centerline{\epsfxsize=3.5truein\epsffile{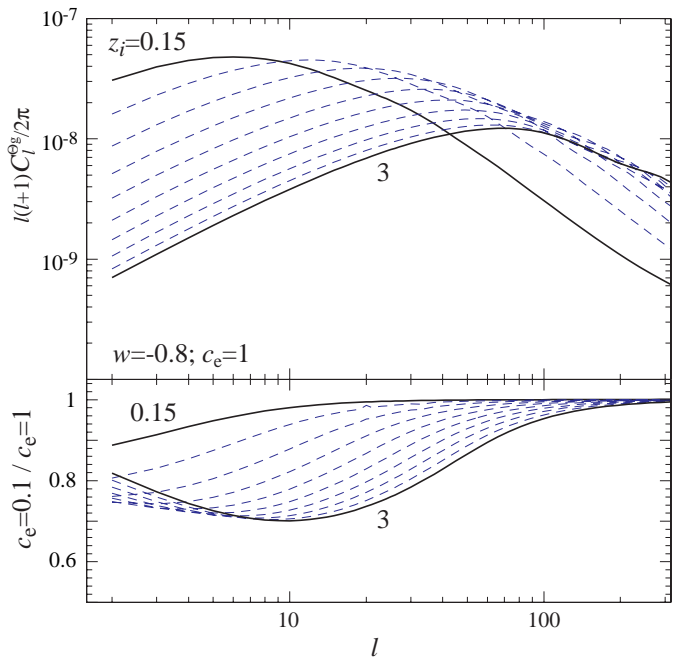}}
\caption{Galaxy-CMB cross correlation through the ISW effect for the $w=-0.8$ model.  
Upper panel: $c_e=1$ cross power
spectra for 10 different galaxy redshift populations of Fig.~\ref{fig:redshift}.
  Lower panel: ratio of the $c_e=0.1$  and $c_e=1$ power spectra.  Note that the
intrinsic 10\% change in the gravitational potential has been amplified by up to 
a factor of 3 
through the sensitivity of the ISW effect to changes in the gravitational potential.}
\label{fig:cross10}
\end{figure}

In Fig.~\ref{fig:cross10}, we show the cross power spectra of these binned galaxy fields
with the ISW effect in the upper panel for $w=-0.8$ and $c_e=1$ and in the lower
panel the ratio of spectra for $c_e=0.1$ and this model.   Note that  due to projection effects,
the correlation in the higher redshift bins peak at higher $\ell$
or  smaller angular scales where the cosmic variance is smaller.

\begin{figure}[t]
\centerline{\epsfxsize=3.5truein\epsffile{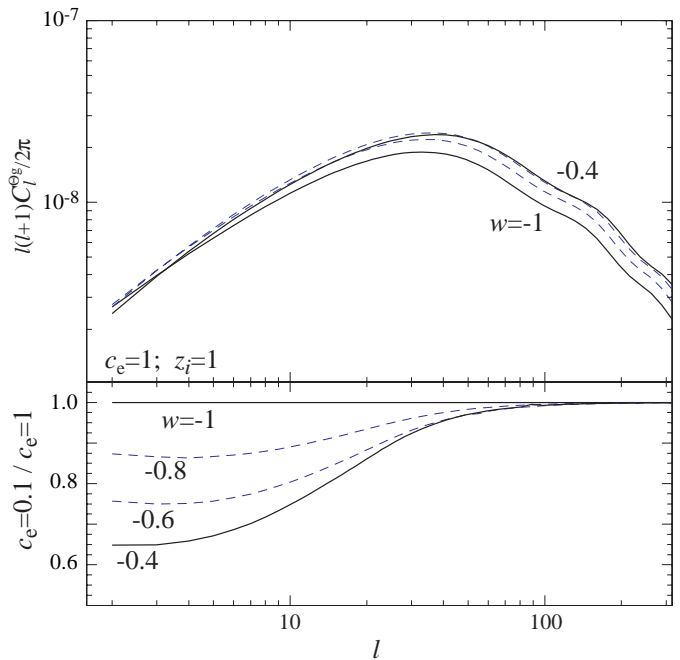}}
\caption{Galaxy-CMB cross correlation as a function of $w$ with the high-$z$ universe
fixed as in Fig.~\ref{fig:phi}.  The cross correlation itself is fairly insensitive to the equation
of state for fixed sound speed ($c_e=1$ upper panel) but becomes increasingly more
sensitive to the sound speed as $w$ increases.}
\label{fig:cross}
\end{figure}

The amplitude of the effect on the cross correlation as parameterized by $c_e$
depends strongly on the background
equation of state.  In Fig.~\ref{fig:cross} we show the cross correlation of the $z=0.75-1$ 
galaxy bin for $w=-1$, $-0.8$, $-0.6$, $-0.4$.   Note that the variations between the
models at $c_e=1$ are relatively small (upper panel) whereas the difference between
$c_e=1$ and $c_e=0.1$ increases as $w$ increases.  The consequence is that
the cross correlation is a relatively insensitive probe of the dark energy equation of
state given the fixed angular diameter distance to recombination and is mainly
useful as a probe of dark energy clustering.  

\begin{figure}[t]
\centerline{\epsfxsize=3.5truein\epsffile{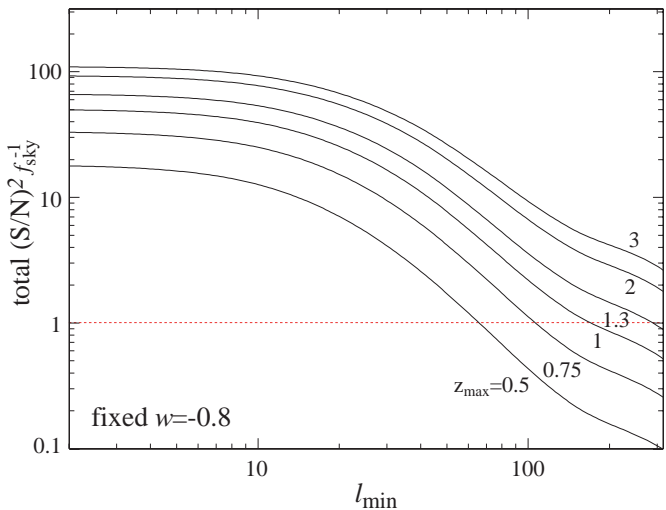}}
\caption{Total significance of the cross correlation detection for galaxies between $z=0$ and
$z_{\rm max}$ and multipoles $\ell_{\rm min} \le \ell \le 1000$.  Half the detection significance
 is supplied
by $z<1$ galaxies and most of the information comes from $10 \le \ell \le 100$.}
\label{fig:totalsn}
\end{figure}

 \section{Clustering Forecasts}
 \label{sec:forecasts}
Interpreting forecasts for the detectability of dark energy clustering requires special care 
for several reasons.    Aside from the special case of
a cosmological constant, no physical model can keep the dark energy smooth near
or above the horizon scale.
It is therefore not physically meaningful to compare a given dark energy model with and 
without dark energy perturbations.
Although the sound speed parameterizes
the physical scale of  the transition to smoothness below the horizon, the amplitude of the
transition depends strongly on the dark energy equation of state $w$ for adiabatic
initial conditions or tracking models which lose their dependence on initial fluctuations.    Since changing the sound speed can shift
the effect outside of an observed range,  differences in models around
a  point in the parameter space do not reveal global
constraints.  

For these reasons,  constraints involving the sound speed
should be translated into  statements as to how smooth the dark energy is out to
a given physical scale. 
Such statements will also retain greater validity outside of the adiabatic and
constant sound speed class of models.

With this in mind let us begin with the usual Fisher approach to parameter estimation forecasts.
Consider the a survey which makes noisy measurements of
a set of fields $x_i$ 
with a noise power spectra of $N_l^{x_i x_j}$.   
Then the total, signal plus noise, power spectra are given by
\begin{equation}
\tilde C_l^{x_i x_j} = C_l^{x_i x_j} + N_l^{x_i x_j}\,.
\end{equation}
Given a parameterization of the signal power spectrum with a set of parameters $p_\alpha$,
the information in the survey on these parameters is quantified  by the Fisher matrix
\begin{equation}
F_{\alpha \beta} =  f_{\rm sky} \sum_l { (2l+1)\Delta l \over 2}
\,{\rm Tr}[ 
{\bf D}_{l\alpha}  \tilde{\bf C}_{l}^{-1}
{\bf D}_{l\beta}  \tilde{\bf C}_{l}^{-1}
]\,,
\label{eqn:Fisher}
\end{equation}
where the sum is over bands of width $\Delta l$ in the power spectra and
$f_{\rm sky}$ is the amount of sky covered by the survey.  
Here 
we have suppressed the $(x_i,x_j)$ indices in a matrix notation and 
\begin{equation}
{\bf D}_{l\alpha} = {\partial {\bf C}_l \over \partial p_\alpha}\,,
\label{eqn:derivs}
\end{equation}
where the derivatives are evaluated at a fiducial model through a finite difference
approximation.
The inverse Fisher matrix approximates the covariance matrix of the
parameters ${\bf C}^p \approx ({\bf F}^{-1})$.  We will take as a fiducial model the 
$w=-0.8$ model of Fig.~\ref{fig:cl}.

For the CMB temperature field $\Theta$,  E-polarization field $E$ and galaxy fields in question, we will assume  white noise power spectra
\begin{eqnarray}
N^{g_i g_i}_l &=&  {1 \over \bar n_{A}^i}\,,\nonumber\\
N^{\Theta \Theta}_l &=& \left( {\Delta_T \over T_{\rm CMB} } \right)^2 e^{\ell(\ell+1)\theta_{\rm FWHM}/8\ln 2}\,, \nonumber\\
N^{E E}_l  &=& \left( {\Delta_P \over T_{\rm CMB} } \right)^2 e^{\ell(\ell+1)\theta_{\rm FWHM}/8\ln 2} \,,
\label{eqn:noise}
\end{eqnarray}
where $n_{A i}$ is the galaxy number density in bin $i$.  All noise cross power
spectra are assumed to vanish.   For the CMB, we
will take for illustrative purposes noise specifications close to the Planck satellite
$\Delta_T = \Delta_P/\sqrt{2} = 40 \mu$K-arcmin and $\theta_{\rm FWHM}=7$ arcmin.
The details of this noise choice are not critical as long as the temperature field
is sample variance limited out to $\ell \sim $ few $\times 10^2$.  
For the galaxies, we use the specifications given in the previous section but again
the number densities are sufficiently high so as to make the shot noise subdominant
for the $\ell$ range in question.

The total signal to noise in the cross power spectra can be quantified through a simple one
parameter family for the spectra
\begin{eqnarray}
C_\ell^{\Theta g_i}& =  &A C_\ell^{\Theta g_i} \Big|_{\rm fid} \,, \nonumber \\
D_\ell^{\Theta g_i} &= & C_\ell^{\Theta g_i} \Big|_{\rm fid}
\end{eqnarray}
such that the Fisher matrix in $A$ gives the significance of the detection given 
a fixed template for the shapes and relative amplitudes
of all the cross power spectra 
\begin{equation}
\left( { S \over N} \right)^2 \Big|_{\rm total} = ({\bf C}^p)^{-1}_{AA} = F_{AA} \,.
\end{equation}
Note that here the noise includes the sample variance of the fields as it represents
the signal-to-noise ratio of the correlation detection.

This total significance is shown in Fig.~\ref{fig:totalsn} and is approximately 100, or
a signal-to-noise ratio of 10 for the full range of $z_{\rm max}< 3$ and $\ell = 2-1000$ for
a full sky survey $f_{\rm sky}=1$.   Even with multiple power spectra from galaxies across
the whole range of the acceleration period, the signal-to-noise remains moderate
due to the sample variance of the single CMB temperature field \cite{Afs04}.

Note that for the $w=-0.8$ fiducial model roughly half of the information comes from galaxies at $z>1$  which
are accessible only to a very deep survey.   (This contribution decreasess somewhat
as $w\rightarrow -1$ leading to a slightly smaller total signal to noise \cite{Afs04}.)
On the other hand, doubling the
information only increases the signal to noise by $\sqrt{2}$.    Furthermore, most of the
information comes from intermediate
multipoles $10 \simlt \ell \simlt 100$ which are accessible to a deep but not necessarily
all-sky survey.   With increasing redshift, the weight moves to higher $\ell$ at higher
$z$ due to projection effects.  

The total signal-to-noise ratio gives a rough quantification of prospects for dark 
energy clustering constraints.   A $S/N$ of 10 implies
that the amplitude of the cross power spectrum can be measured to about $10\%$ from
the intermediate multipoles.  Since the ISW effect is sensitive
to the change in the potential, it gains a factor of $\sim 3$ enhancement in sensitivity to
potential variations.   Thus roughly the cross correlation
can detect a $10\%$ variation in $\Phi$ as in the $w=-0.8$ model 
at $\sim 3\sigma$.  This $\ell$ range implies that this ability applies to
a transition scale of  $1\%-10\%$ of the horizon. 
 
This order of magnitude estimate is borne out by a more quantitative treatment.  
Here we employ the 7
cosmological parameters that determine the fiducial model in Fig.~\ref{fig:cl} 
as Fisher parameters.
To account for the dark energy clustering we add the sound speed as a parameter.  
Because the effect of the sound speed is only apparent for variations that span an order
of magnitude in $c_e$, we take this parameter to be
\begin{equation}
p_{c_e} = { \log_{10} c_e \over \Delta \log_{10} c_e}\,,
\end{equation}
where $\Delta\log_{10} c_e$ is the step away from $c_e=1$ used to define the parameter derivative in 
Eq.~(\ref{eqn:derivs}).   This normalization factor reflects the fact that the Fisher
errors should be interpreted as the significance of the observed difference between the
actual models used to define the derivative
\begin{equation}
\left( { S \over N} \right)^2 \Big|_{c_e} = ({\bf C}^p)^{-1}_{p_{c_e} p_{c_e}}  \,.
\end{equation}
accounting for parameter degeneracies.

\begin{figure}[t]
\centerline{\epsfxsize=3.5truein\epsffile{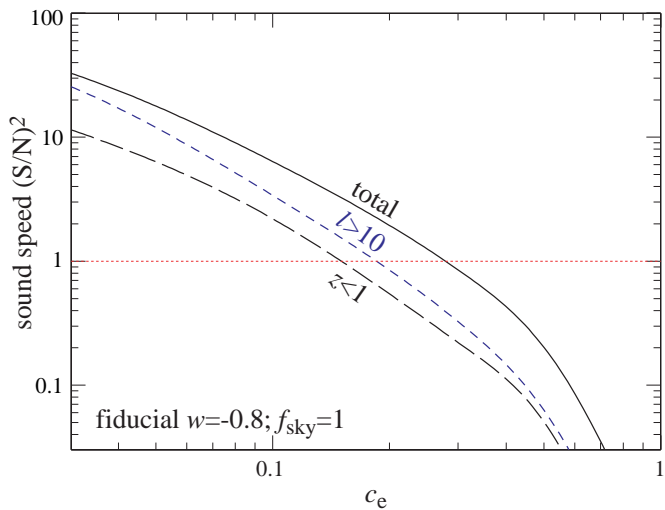}}
\caption{Significance of the separation between a given $c_e$ model and 
a quintessence model of $c_e=1$  given an all sky survey of galaxies and allowing
for parameter degeneracies around a $w=-0.8$ model. 
Note that in comparison with the total detection significance, dark energy clustering model
separation requires lower multipoles and utilizes more high-$z$ galaxies as $c_e \rightarrow 1$
and the effects are confined to the largest scales.}
\label{fig:signif}
\end{figure}

To these cosmological parameters, we supplement 5 additional ``nuisance" parameters per galaxy 
redshift
bin that define the galaxy bias.   These parameters, given explicitly in \cite{HuJai03},
define the occupation of galaxies in dark matter halos.   We then have 
a total of 58 parameters. 
Though marginalization of these extra parameters can in general
degrade constraints due to parameter degeneracies, in this case
the information in the  auto power spectra of the CMB and galaxies suffices to make the
constraints quite comparable to those from employing the cross power spectrum only but
with the other parameters fixed.  The addition of the galaxy auto correlation power spectrum
actually enhances the significance of the separation between $c_e=1$ and $c_e\ll 1$ models
and begins to play a role at $c_e \simlt 0.1$. 
Similar to massive neutrinos, for very low
sound speeds, the transition occurs on scales where the galaxy power spectra are well measured 
and the ISW effect is negligible.  However there may be subtleties in the $c_e\ll 1$ regime
involved with galaxy bias
as to whether the galaxies trace the total density perturbation or that of the dark matter.

We show the significance of the separation between models with different
sound speeds in Fig.~\ref{fig:signif} including marginalization over $w$ and other parameters.  
Recall that for the $w=-0.8$ fiducial model
shown here the amplitude of the effect is $\sim 10\%$ in the gravitational potential.
The model with $c_e=0.1$ is distinguished at a $S/N \approx 2.5$ for $f_{\rm sky}=1$.  Note however that the
significance drops off rapidly as $c_e \rightarrow 1$ as the changes become confined
to the low multipoles.  Thus one would draw an incorrect inference if one employed 
$\log_{10} c_e$ directly as a Fisher parameter and let the derivative be approximated by
finite difference with  $\Delta \log_{10} c_e \rightarrow 0$.

\begin{figure}[t]
\centerline{\epsfxsize=3.5truein\epsffile{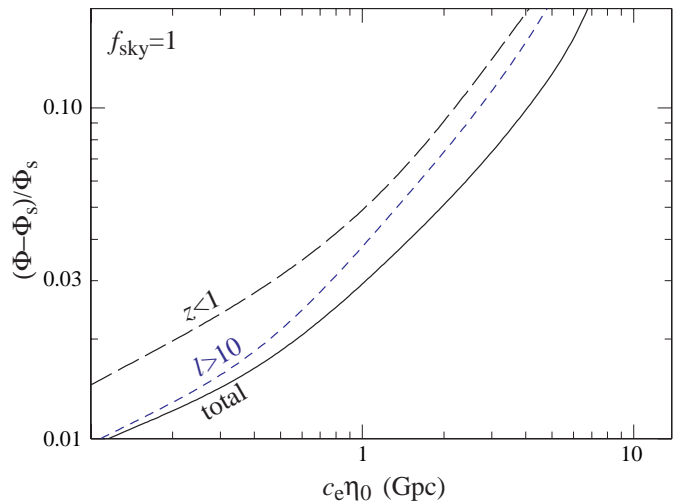}}
\caption{Conversion of the sound speed constraints into more robust
dark energy smoothness constraints.  Shown are the projected $1\sigma$ errors
on the change in the gravitational potential due to dark energy clustering as a function
of scale.  The galaxy-ISW correlation can ultimately constrain the smoothness of
the dark energy at the 1Gpc scale to $\sim 3$\% in the potential.}
\label{fig:cluster}
\end{figure}

Utilization of the ISW-galaxy correlation for dark energy clustering does shift the optimal redshift and multipole range to higher redshifts and lower multipoles compared with a simple detection
significance
criteria.  Because the effects are confined
to large physical scales, they can only be observed out to a certain 
maximum $\ell$ which increases with redshift due to projection effects (see Fig.~\ref{fig:cross10}).
The effect of losing the lowest multipoles due to survey boundaries or systematic effects
or the highest redshifts due to depth and photometric redshift errors
is thus more substantial for distinguishing between clustering models (see Fig.~\ref{fig:signif}).

These results remain robust when varying the fiducial model in $w$ if interpreted as 
the significance of detecting a given variation in the gravitational potential due to dark
energy clustering on scales corresponding to $ c_e \eta_0 \approx 14 c_e$ Gpc.  We show
this rescaling of Fig.~\ref{fig:cluster} to $S/N=1$ for effective $1\sigma$ errors on the change
in the gravitational potential due to dark energy clustering out to a given scale.
The caveat is
that as $w \rightarrow -1$ one would predict no change out through the horizon
in an adiabatic model.
These results are also robust to changes in the halo model for galaxy bias, variations in
the galaxy source density, and improvements
in photometric redshift accuracy.

\section{Discussion}

The ISW effect provides a unique probe of the dynamics of the dark energy through its high
sensitivity to the smoothness of the dark energy.
 Although the significance of its detection through
cross correlation with galaxy surveys or gravitational lensing statistics will never
exceed $\sim 10 \sigma$ due to the cosmic variance of the CMB temperature field \cite{Afs04},
it is likely to be the best means of probing the clustering dynamics since their effects
appear only on the largest scales.   

A $10 \sigma$ detection of the correlation will allow its measurement to $\sim 10\%$ and
hence yield the ability to limit $\sim 3\%$ variations in the gravitational potential
due to the dark energy clustering if they appear on the Gpc scale.    In the adiabatic
model, this would allow for a test of the quintessence, or canonical single scalar field, hypothesis
if $|1+w| \simgt 0.05$.   For distinguishing clustering models, as opposed to a simple
detection of the correlation, it is important to measure correlation at the largest angles
out to $z \sim 2$. That will require a deep nearly all-sky survey such as LSST \cite{LSST}.  
Furthermore the galaxy clustering signal on very large angles is itself small and controlling
systematics  in the galaxy surveys will pose a significant challenge (e.g.
\cite{MadEfsSut90,Scretal04b}). 

Finally, beyond the adiabatic model, clustering in the dark energy may have strong effects
even if $w\rightarrow -1$.  For example, in the isocurvature models designed to suppress the quadrupole
\cite{MorTak03,GorHu04}, the change in the gravitational potential is about
6 times that of the 
adiabatic model.  Such models are potentially testable with CMB-galaxy cross correlations
in the right redshift range.  Likewise explanations for the acceleration that involve modifications
to gravity in place of the dark energy may predict a different CMB-galaxy correlation on the
largest scales.   The uniqueness of this probe of the acceleration of the expansion 
should motivate future studies despite challenges facing its exploitation.

\smallskip
\noindent{\it Acknowledgments:}  We thank J. Carlstrom, L. Pogosian, C. Gordon, and A. Stebbins
for useful discussions.
WH was supported by the DOE and the Packard Foundation. RS was partially supported by NSF CAREER award AST99 84924 and 
NSF ITR ACI-0121671.
A portion of this work was carried out at the KICP under NSF PHY-0114422.

\end{document}